\documentstyle[12pt,epsf,epsfig]{article}
\def\be{\begin{equation}}
\def\ee{\end{equation}}
\def\ba{\begin{eqnarray}}
\def\ea{\end{eqnarray}}

\def\bdm{\begin{displaymath}}
\def\edm{\end{displaymath}}
\def\la{~\mbox{\raisebox{-.6ex}{$\stackrel{<}{\sim}$}}~}

\def\bq{\begin{quote}}
\def\eq{\end{quote}}

 at 10truept
\def\la{~\mbox{\raisebox{-.6ex}{$\stackrel{<}{\sim}$}}~}

\newcommand{\beq}{\begin{equation}}
\newcommand{\eeq}{\end{equation}}
\newcommand{\bea}{\begin{eqnarray}}
\newcommand{\eea}{\end{eqnarray}}
\newcommand{\beqa}{\begin{eqnarray}}
\newcommand{\eeqa}{\end{eqnarray}}

\def\ltap{\ \raise.3ex\hbox{$<$\kern-.75em\lower1ex\hbox{$\sim$}}\ }
\def\gtap{\ \raise.3ex\hbox{$>$\kern-.75em\lower1ex\hbox{$\sim$}}\ }
\def\gl{\ \raise.5ex\hbox{$>$}\kern-.8em\lower.5ex\hbox{$<$}\ }
\def\roughly#1{\raise.3ex\hbox{$#1$\kern-.75em\lower1ex\hbox{$\sim$}}}

\begin{document}

\thispagestyle{empty}
\begin{flushright}
March 2012
\end{flushright}
\vspace*{1.5cm}
\begin{center}
{\Large \bf Cutoffs, Stretched Horizons and Black Hole Radiators}

\vspace*{1.5cm} {\large Nemanja Kaloper\footnote{\tt
kaloper@physics.ucdavis.edu} 
\\
\vspace{.8cm}  
Department of Physics, University of
California, Davis, CA 95616, USA}\\

\vspace{1.7cm} ABSTRACT
\end{center}
We argue that if the UV cutoff of an effective field theory with many low energy degrees if freedom is of the order, or below, the scale of the stretched horizon in a black hole background, 
which in turn is significantly lower than the Planck scale, the black hole radiance
rate may not be enhanced by the emission of all the light IR modes.
Instead, there may be additional suppressions hidden in the UV completion of the field theory, which really control which light modes can be emitted by the black hole. 
It could turn out that many degrees of freedom cannot be efficiently emitted by the black hole, and so the radiance rate may be much smaller than its estimate based on the counting of the light IR degrees of freedom.
If we apply this argument to the RS2 brane world, it implies that the emission rates of the low energy CFT modes will be dramatically suppressed:
its UV completion is given by the bulk gravity on $AdS_5 \times S^5$, and the only bulk modes that could be emitted by a black hole are the 4D $s$-waves of bulk modes with small $5D$ momentum, or equivalently, small $4D$ masses. Further, their emission is suppressed by bulk warping, which lowers the radiation rate much below the IR estimate, yielding a radiation 
flux $\sim (T_{BH} L)^2 {\cal L}_{hawking} \sim (T_{BH}/M_{Pl})^2 N {\cal L}_{hawking}$, where ${\cal L}_{hawking}$ is the Hawking radiation rate of a single light species. This
follows directly from low CFT cutoff $\mu \sim L^{-1} \ll M_{Pl}$, a large number of 
modes $N \gg 1$ and the fact that $4D$ gravity
in RS2 is induced, $M_{Pl}^2 \simeq N \mu^2$.

\vfill \setcounter{page}{0} \setcounter{footnote}{0}
\newpage

\section{Introduction}

One of the greatest surprises in our understanding of black holes came when Hawking showed that they are thermal objects, emitting quantum radiation at a universal rate controlled by the black hole mass \cite{hawking}. This confirmed Bekenstein's argument \cite{bekenstein} that a black hole is a statistical system with an entropy given by its area. Further development of black hole thermodynamics led to many puzzles and confusions. The effort to resolve them led to  the holographic principle \cite{holography}, which posits that if quantum gravity is unitary, black hole degrees of freedom must be packed on the horizon area, rather that inside the black hole's volume. 

Formulating these ideas in effective field theory is complicated by the divergences which appear when field theory states are brought to black hole horizons. If one insists on trying to describe near-horizon quantum states within field theory, one must regulate the divergences. A simple regulator is a stiff cutoff as in the brick wall method \cite{thooft}. Better yet, one can use the concept of the stretched horizon \cite{lenny}, which has similar effects while preserving local covariance. With it in place, cutting out the divergent, `trans-planckian' region near the horizon, the field theory quantities are insulated from the black hole horizon divergences. The location of the stretched horizon, however, should be picked carefully, in order to reproduce the information content of a black hole accessible to field theory probes. 
 
In `simple' examples of quantum field theories, with a few degrees of freedom, weak couplings outside of a black hole, and the IR modes unchanged all the way
to the Planck scale, the stretched horizon is a timelike surface `hovering' just outside of the event horizon, at a Planckian proper distance $\sim \ell_{Pl}$ from it \cite{lenny}. It is there where the thermodynamical entropy in the infalling `vacuum' state equals the Bekenstein-Hawking entropy of a black hole. 
On this surface the blueshifted black hole temperature is of the order of the Planck mass $M_{Pl}$, a natural cutoff of the quantum field theory describing a black hole. Clearly, this is much higher than the black hole temperature measured at infnity. Thus one may take this to be the `physical' hot surface of a black hole from where the hot Hawking quanta originate, and eventually propagate out to infinity while getting redshifted down to fit the black body spectrum with temperature $T_{BH} \sim M_{Pl}^2/M$ seen by a distant observer, where 
$M_{Pl}^2 = 1/\ell_{Pl}^2 = (8\pi G_N)^{-1}$. The stretched horizon is the Hawking `radiator', in effect coarse graining the detailed physical mechanisms yielding the radiance. 

When one considers a field theory outside of a black hole with  $N\gg 1$ degrees of freedom the story is more complicated. Guided 
by the equivalence principle, one may guess that since all species exterior to the black hole gravitate in the same way,  thermalizing with one means thermalizing with all. Even if the field theory is strongly coupled, 
it is still subject to standard thermodynamics \cite{thermo}, and as long as its modes are light they can carry out radiation, as much as grey body factors permit. Since the modes originate near the horizon, to be sure if they are created one would need  to push such an equilibrated state inward to the (fixed) Planckian distance from the black hole. But with many modes, this is a no-go: the naive state's entropy would have overshot the Bekenstein-Hawking entropy by a large factor, $ \la N$. So using the stretched horizon far from the black hole, one will obtain the correct scaling of the entropy with the horizon area. But one cannot be sure just how many of these modes will be emitted, because one may be missing any suppression (or enhancement!) processes in the excised region, where classical geometry is still well defined.

The point is that one must be careful with extrapolating the notion of `thermalization' of a quantum field theory with the black hole. The caveat is that while near the black hole all the modes will be thermally excited, not all may be able to flee far from it. This shows up already in weakly coupled quantum field theories with few species and Planckian cutoff. In this case, the thermal state of a black hole is coded by picking the quantum state of the field theory to be the infalling vacuum, in order to ensure the regularity of the event horizon after quantum backreaction is included. This state is defined by the boundary conditions for the mode functions at the horizon. In terms of the `out' modes this state turns out to be occupied with excitations of all modes that follow the thermal distribution near the horizon. However, not all these modes (for example, excitations with arbitrarily large spin) can be efficiently emitted to infinity. Most of them are prevented from escaping by the centrifugal barrier, proportional to their spin,
which reduces the emissivity by inducing a large gray body factor. Similarly, if there are modes heavier than the black hole temperature at infinity, while they are excited near the horizon, they cannot escape because of the mass barrier, unless their kinetic energy exceeds their gravitational potential energy in the black hole background.
Since all this occurs at, or well below the UV cutoff, we can completely consistently determine the black hole radiance rate using the field theory below the cutoff.
However, if the UV cutoff is low, and the stretched horizon far, while the stretched horizon may correctly reproduce black hole entropy, the theory there and farther from the black hole may not be able to resolve any effects that may alter the radiance that happen inside between the stretched horizon and the black hole. Shedding some new light on this issue is the central point of this paper.

In what follows we will argue that to deduce the correct radiation rate from a black hole whose exterior supports an effective field theory with a low UV cutoff and many species of light modes at low energies, we may need to consider the UV completion of the field theory. Specifically, if
the stretched horizon is parametrically much farther than the Planck distance $\ell_{Pl}$ from the event horizon, and if it is above, or close to the UV cutoff\footnote{Which is therefore also much below the Planck scale.} one should first UV-complete the field theory (close) to the Planck scale, and move it inward, beyond the stretched horizon set by the IR contents. This will not change the field theory entropy count of the modes excited by the black hole. But it may  turn out that most of the UV modes cannot be efficiently emitted by a black hole, so that  the naive radiance rate estimated by including all the IR degrees of freedom may be drastically reduced. 
This argument should be useful for approaching the problem of Hawking radiance in models which contain lots of composite particles in the IR, and in the framework of induced gravity \cite{inducedgr}. The situation where the stretched horizon is well below the UV cutoff which in turn is below the Planck scale is ambiguous, and we won't have much to say about it in detail, although we will suggest an example to further test what happens in this case.

An example where the stretched horizon determined by the counting of the many light modes of a low energy effective field theory is comparable to the vacuum UV cutoff of the low energy theory is the
RS2 brane world \cite{RS2}. The  IR field theory is a $SU({\cal N})$ super-YM CFT in strong coupling regime, with many degrees of freedom in the large ${\cal N}$ limit, $N \sim {\cal N}^2$. It is cut off at a scale $\mu \sim L^{-1}$, where $L$ is the $AdS_5$ radius, and it couples to $4D$ gravity \cite{rsholo,gubser}, which is however {\it induced}, having the $4D$ Planck scale given by $M^2_{Pl} \sim N \mu^2$ \cite{induced}. The entropy matching, taking all the $N$ degrees of freedom of the CFT to be thermalized with a black hole sitting on the RS2 brane \cite{efk,takahiro}, requires that the multispecies stretched horizon is at the distance $\sim \sqrt{N} \ell_{Pl} \sim L$ from the event horizon, which is to say, precisely at the CFT cutoff. So the CFT at this distance from the hole is much hotter than at infinity, with the temperature given by the CFT cutoff $T = T_0/\sqrt{g_{00}(r_S)} \sim (\sqrt{N} \ell_{Pl})^{-1} \sim \mu$. Closer to the horizon, where the Hawking radiation originates, the CFT doesn't even exist. To see how much energy is radiated to infinity, we can use the UV completion to go beyond the cutoff $\mu \sim L^{-1}$. 
The UV completion of RS2 above the cutoff $L^{-1}$ is the weakly coupled bulk gravity (actually, strictly speaking, full string theory on $AdS_5 \times S^5$, in weak coupling limit), and it exists close to the horizon. In this description, the only bulk modes that could be emitted by a black hole are the $s$-waves of normalizable 
bulk zero modes with small $4D$ masses (which are constant on $S^5$). 
The probability of emitting these modes is suppressed by bulk warping. Perturbatively, they are competing with the RS2 volcano potential. The volcano suppresses the emission rate much below the IR estimate, yielding a radiation 
flux $\sim (T_{BH} L)^2 {\cal L}_{hawking} \sim (T_{BH}/M_{Pl})^2 N {\cal L}_{hawking}$, where ${\cal L}_{hawking}$ is the Hawking radiation rate of a single species. 
This is analogous to what occurs for black holes in flat large dimensions \cite{savas}: there the centrifugal barrier induces grey body factors for bulk modes 
which lower the radiation rate much below the $4D$ IR estimates \cite{robertoco}. Here, it is the bulk potential barrier induced by warping that yields the extra suppression. In consequence, by energy conservation, the emission rates of the low energy CFT modes will be very suppressed.

If the flat space UV cutoff is much higher than the temperature on the stretched horizon, the description is more ambiguous. On the one hand, perturbative field theory arguments suggest that IR theory ought to remain reliable, and its low energy modes should emerge on the stretched horizon even if it does not extend all the way down to the fundamental Planck scale. So if the number of IR modes below the cutoff is still very large, a black hole might emit many more modes,  if they all couple to gravity universally and do not suffer additional suppressions. On the other hand, this case is very different from the previous situation where the UV cutoff of the effective field theory is low. One may argue \cite{gia} that with many species present the flat space cutoff is incorrect \cite{gia} since it ignores strong gravitational effects\footnote{These effects are not related to the strong coupling effects in the field theory.} which are known to dominate in the UV \cite{thooftdominance}, and one needs to include their effects when renormalizing the theory in a black hole background, which may
lower the UV cutoff. We can't say much about this, except to point out that this could be studied in a setup which is  a hybrid between an RS2 braneworld and a {\it negative} DGP-like term localized on the brane, 
with a large coefficient. In this note we will not address this in detail, leaving further exploration for the future.

\section{A Review of Black Hole Radiance}

Let us start by briefly reviewing the derivation of Hawking radiation from a black hole. Following \cite{hawking}, perhaps the simplest setup to consider is the dynamical black hole
geometry, where the black hole forms in a collapse of a null shell. This is an approximation to a realistic black hole formation from collapse of a cloud of matter. The calculation involves setting up the classical background, and then treating quantum field theory on it using perturbation theory in the interaction
picture. The Hamiltonian and the mode functions governing the system will all jump across the shell, whereas the quantum state which the system occupied before the transition, i.e. the flat space vacuum, will remain the same\footnote{As long as the theory used to describe it remains under calculational control, ie this all occurs below the field theory cutoff.}. It will, however, not be the eigenstate of the new Hamiltonian, and this is precisely why Hawking radiation arises. Note, that the  transition of the quantum state from the non-radiating vacuum to a radiation-filled state is continuous, albeit extremely rapid.  

\begin{figure}[hbt]
\begin{center}
\includegraphics[width=0.5\textwidth]{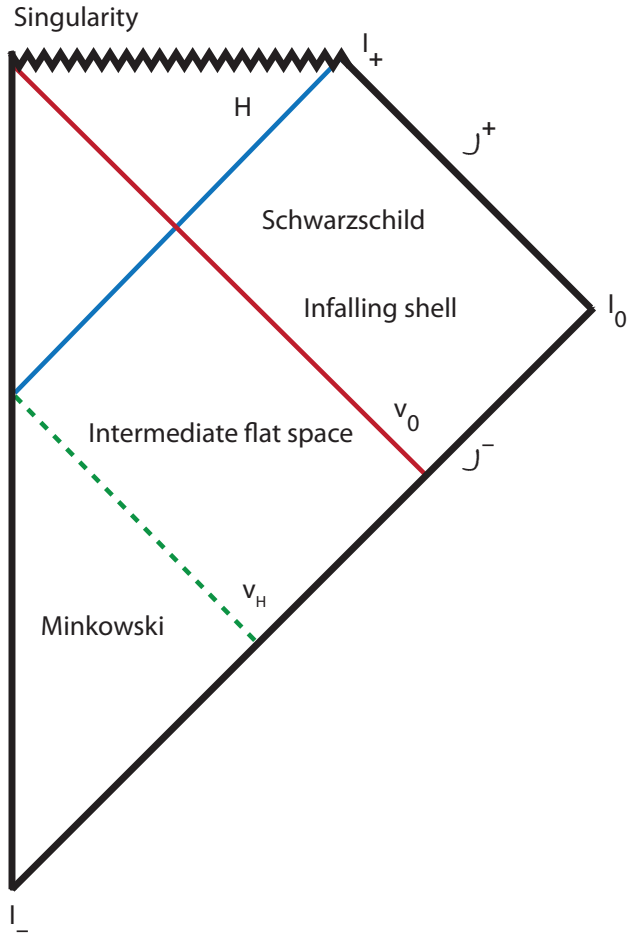}
\caption{Causal diagram of a black hole formed by a burst of radiation along $v_0$}
\label{one}
\end{center}
\end{figure}
The background geometry is described by a variant of the Vaidya metric,
\be
ds^2 = - \Bigl(1-\frac{r_0(v)}{r}\Bigr) dv^2 + 2dvdr + r^2 d\Omega_2 \, ,
\label{vaidya}
\ee
where $r_0(v) = r_0 \theta(v-v_0)$, $\theta$ is the step function, $r_0 = 2 G_N M$ is the horizon of the final black hole, and $v_0$ is the ingoing 
null lightcone corresponding to the imploding shell worldvolume. The coordinate $v$ is the outgoing null coordinate, related to the standard time by the transformation $v_M=t+r$ in the flat regions, and $v_S = t + r + r_0\ln\Bigl(r/r_0 - 1\Bigr)$ in the Schwarzschild region. Accordingly, we can also define the ingoing null coordinate $u$ which measures time along an ingoing null ray $v = {\rm const.}$ by $u_M = t - r$ in the flat regions and
$u_S = t - r - r_0\ln\Bigl(r/r_0 - 1\Bigr)$ in the Schwarzschild region. The causal diagram of this geometry is given in Figure 1. 
The region of spacetime inside the shell at $v_0$ is just the Minkowski space, however in the part above the past lightcone $v_H$ it is divided into two 
causally separated regions by the black hole event horizon. The inside is trapped, while the outside is normal, like its parent Minkowski region in the past. The location of the boundary lightcone $v_H$ is given by the first instant of the emergence of the event horizon, which is dynamical in the intermediate region, and starts at zero radius at the time $t=v_H$, growing to $r=r_0$ at the time $t_0 = v_0 - r_0$. Since the horizon itself is a null
surface, and it starts with zero size, its location in the intermediate flat region is determined by the equation $t = r + v_H$, yielding $t_0 = r_0+v_H$ at $v_0$. Combining these equations gives $v_H  = v_0 - 2r_0$. Lastly, since flat and Schwarzschild regions are matched along the imploding
null shell at $v_0$ with a nonzero stress energy $T^\mu{}_\nu \propto r_0 \delta(v-v_0)$, and since the matching conditions require that the spheres on either side of the null shell have equal areas, the $u$ coordinate jumps across the 
shell. This follows from $v_M - u_M = 2r$ and $v_S - u_S = 2r + 2r_0 \ln\Bigl(r/r_0-1\Bigr)$. Imposing that the area is the same on either side of $v_0$ yields 
\be
u_S = u_M - 2r_0 \ln(\frac{v_H - u_M}{2r_0}) \, . 
\label{matching}
\ee
The difference is the boost factor taking into account the presence of the horizon in the intermediate 
flat region, and allows one to match together the Minkowski and Schwarzschild charts in the atlas of Figure 1. The coordinates $u_M$ and $u_S$ can
be extended up above $v_H$ and down below $v_0$, respectively, and matched together using this formula. The boost is controlled by
$v_H$, set by the time when the horizon first arises. Afterwards, the constant time surfaces are all affixed to the intersection of $H$ and $v_H$ surfaces in the Figure 1, bending upward and never crossing the horizon $H$ nor the past lightcone $v_H$.  

Let us now see how the black hole is modeled in field theory. Consider a weakly coupled field theory in this background, for example a minimally coupled massless scalar field. Its field equation,
$\nabla^2 \Phi = 0$, can be expanded in the eigenmodes of the asymptotic symmetries of the geometry, namely rotational group
and time translation generators. So, substituting $\phi = e^{\mp i \omega t} Y_{lm}(\phi,\theta) {\varphi_\omega(r)}/{r}$, and introducing the tortoise coordinate
$r_* = r + r_0 \ln\Bigl(r/r_0 - 1\Bigr)$, the radial modes solve the equations
\ba
&& \varphi_\omega'' + \Bigl(\omega^2 - \frac{l(l+1)}{r^2}\Bigr) \varphi_\omega = 0 \, ,    
 \, ~~~~~~~~~~~~~~~~~~~~~~~~~~~~~~~~~~~ {\rm in~ Minkowski}\, ,  \nonumber \\
&& \ddot \varphi_\omega + \Bigl(\omega^2 - \frac{l(l+1)}{r^2}(1-\frac{r_0}{r}) - \frac{r_0}{r^3} (1-\frac{r_0}{r})\Bigr) \varphi_\omega = 0 \, , 
 ~~~~~~~ {\rm in~ Schwarzschild} \, . \label{radial}
\ea
The primes are derivatives with respect to $r$, while the dots are derivatives with respect to the tortoise coordinate $r_*$. 
These equations show that generic modes have different support near and far from the black hole, thanks to the centrifugal barrier
$\propto l(l+1)$, which suppresses very strongly the wavefunction of $l \ne 0$ far from the black hole. Thus the dominant channel for
black hole radiance must be the $s$-wave, $l=0$ \cite{robertoco}. This mode is still suppressed on the way out of the black hole by the potential 
barrier $V_B = \frac{r_0}{r^3}(1-r_0/r)$, but this suppression is much weaker than the one from the centrifugal barrier. The $l \ne 0$ modes serve as the reservoir of black hole entropy and energy, which can only escape when the local interactions convert it into $s$-waves.

To determine the Hawking spectrum to the leading order one can work in perturbation theory, first ignoring
the barriers altogether, and determining the energy flux in two steps. The first step is to find the particle contents 
of the quantum state near the black hole, to the far left of the potential barrier in the tortoise coordinate. In this limit, the black 
hole is approximated by the Rindler geometry, and so the particles that are in the thermal bath near it can be referred to as the 
Unruh radiation \cite{unruh}. The second step is to find what fraction of this near-horizon heat
can escape to infinity through the barrier $V_B$ where it can be harnessed away. This is the `actual' Hawking radiation, which carries energy and entropy out of the black hole.

Let us start with determining the Unruh radiation of the $s$-wave. In this case one can approximate (\ref{radial}) with
\ba
&& \varphi_\omega'' + \omega^2 \varphi_\omega = 0 \, ,    
~~~~~~~ {\rm in~ Minkowski}\, ,  \nonumber \\
&& \ddot \varphi_\omega + \omega^2 \varphi_\omega = 0 \, , 
 ~~~~~~~ {\rm in~ Schwarzschild} \, . \label{radial2}
\ea
The mode functions $\Phi_\omega = \varphi_\omega/r$ are, after restoring the time dependence, and using null coordinates $v$ and $u$,
\be
\Phi_\omega(r)= \frac{1}{r} \cases{A e^{-i\omega u_M} + B e^{-i\omega v_M} \, , ~~~~~~ {\rm in~Minkowski} \, ,\cr 
\bar A e^{-i\omega u_S} \, + \, \bar  B e^{-i\omega v_S}  \, , ~~~~~~ {\rm in~Schwarzschild} \, .}
\label{modes}
\ee
The ``in" modes, defining the Fock space of the initial Minkowski region, are positive frequency ingoing and outgoing modes, 
$A e^{-i\omega u_M}/r$ and $B e^{-i\omega v_M}/r$, respectively. The ``out" modes, which define the Fock space of an exterior observer
on ${\cal J}^+$, and the near-horizon region $H$, and therefore reside on $t = {\rm const.} \gg v_H$ slices in the Schwarzschild chart are 
$\bar A e^{-i\omega u_S}/r$ and $\bar B e^{-i\omega v_S}/r$. The ``in" and ``out" modes are nontrivially related, because of the 
blueshifts in the geometry due to the infalling shell at $v_0$.  To match them, note first that
an outgoing positive frequency ``out" mode in the intermediate flat region is $
\frac{\bar A}{r} e^{-i \omega u_S} = \frac{\bar A}{r} \Bigl(\frac{v_H - u_M}{2r_0}\Bigr)^{2i r_0 \omega} e^{-i \omega u_M} $
after substituting (\ref{matching}). At times $t < v_H$, since the ``out" modes should evolve from completely regular initial data, we remove the singularity in the mode functions at $r=0$ by adding the ingoing mode to 
$\Phi_\omega^{``out"}{}_{outgoing}$, with the support in only the initial Minkowski region. This means, we impose reflective
boundary conditions at $r=0$. So, the regular continuation of 
the outgoing ``out" positive frequency mode is \cite{patatino}
\be
\Phi_\omega^{``out"}{}_{outgoing} = \frac{\bar A}{r} \Bigl[\Bigl(\frac{v_H - u_M}{2r_0}\Bigr)^{2i r_0 \omega} e^{-i \omega u_M} 
- \Bigl(\frac{v_H - v}{2r_0}\Bigr)^{2i r_0 \omega} e^{-i \omega v} \theta(v_H - v) \Bigr] \, .
\label{outmodepast}
\ee
Here we dropped the subscript on $v$ since it doesn't jump across the shell $v_0$. Substituting $u_M = t-r$ and $v = t+r$ here, taking $v < v_H$, we find that in the limit $r \rightarrow 0$, the mode reduces to
$\Phi_\omega^{``out"}{}_{outgoing} \rightarrow 2i \omega (\frac{t+r_0}{2r_0})^{2i r_0 \omega} e^{-i\omega t - 2r_0 \omega \pi}$,  and it is nonsingular. We can similarly extend the formula for the ``out" ingoing modes. Finally, we pick the normalization constants $\bar A =1/[2\omega(2\pi)^3]^{-1/2}$
to satisfy the standard mode orthogonality relations, $\langle \Phi_\omega^{``out"}{}_{outgoing}, \Phi_{\omega'}^{``out"}{}_{outgoing} \rangle = \delta(\omega-\omega')$, and analogous ones for other modes, in the ``in" and ``out" zones. The bottomline is that the regular
outgoing ``out" modes must have nonzero overlap with ingoing ``in" modes, as is clear from the presence of the second term in (\ref{outmodepast}). However these cross terms involve
{\it both positive and negative frequencies}. This follows from the power law prefactor, which enforces the mixing between frequencies 
of different signs.

These results in fact directly apply to the $l \ne 0$ modes as well. It is clear from (\ref{radial}) that in the Schwarzschild region, these modes obey the same equation as $s$-waves near the horizon, since the centrifugal barrier is negligible because of the redshift factor $\propto (1-r_0/r)$
in (\ref{radial}). The centrifugal barrier in the Minkowski region cannot be neglected near the origin, but its role there is to merely
prevent the Minkowski region radial wavefunctions from blowing up, replacing the radial modes by $J_{\sqrt{l(l+1)+1/4}}(\omega r)/\sqrt{\omega r} $ Bessel functions. However, the Schwarzschild radial modes are completely unaffected by this, and still need to be regulated once they are pulled back to the Minkowski region, by the addition of a continuation of the Schwarzschild region ingoing mode, 
just like the $s$-waves in (\ref{outmodepast}). Because the Bessel functions  $J_{\sqrt{l(l+1)+1/4}}(\omega r)/\sqrt{\omega r} $, representing 
Minkowski region radial modes, reduce to precisely the usual flat space ingoing modes as $r\rightarrow \infty$, where we compute the mode overlaps
$\langle \Phi_\omega^{``out"}{}_{outgoing}, \Phi_{\omega'}^{``in"}{}_{outgoing} \rangle$, the Minkowski region short distance modification is insignificant and  the $l\ne 0$ modes will behave exactly as the $s$-waves, up to an irrelevant phase. 

\begin{figure}[hbt]
\begin{center}
\includegraphics[width=0.55\textwidth]{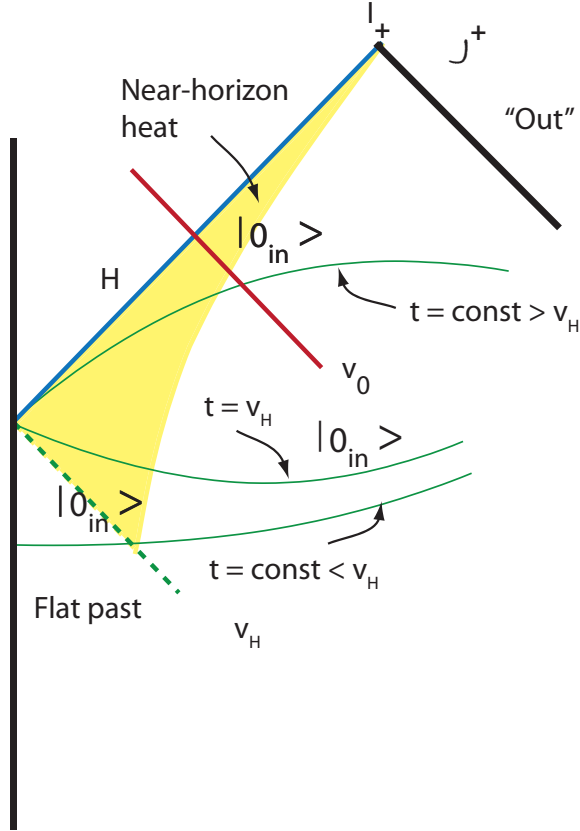}
\caption{Near-horizon heat}
\label{two}
\end{center}
\end{figure}
Next, let us consider the quantum state of the system (\ref{vaidya}). Since the system is prepared as an initially infinite section of a flat space, 
we can take the quantum state inside the shell to be the ``in" vacuum $|0_{in} \rangle$. 
This state is defined as the state annihilated by all
the positive frequency ``in" operators, i.e. $\hat \Phi_{\omega>0}^{``in"}  \, | 0_{in} \rangle = 0$. These operators are the 
Fourier transforms of the field $\Phi$, satisfying $\hat \Phi_{\omega>0}^{``in"} = \langle \Phi_{\omega>0}^{``in"}{}_{~outgoing,ingoing}~~ , \Phi \rangle$. 
The ``in" outgoing and ingoing wavefunctions, which play the role of the Fourier kernel, are given by $e^{-i\omega u_M}/[2\omega(2\pi)^3]^{1/2}$ and $e^{-i\omega v}/[2\omega(2\pi)^3]^{1/2}$, respectively. As this state evolves in time, propagating from one $t=const.$ slice to another, as depicted in Figure 2, in the interaction picture, it is not altered by the imploding shell. However, at late times $ t \gg v_H$, the field operators are all evolved to $\hat \Phi_\omega^{``out"}$. As we noted above, these are nontrivial mixtures of $\hat \Phi_\omega^{``in"}$, by virtue of Eq. (\ref{outmodepast}) and related formulas for ingoing modes. 
These formulas enable one to compute the mixing, by projecting $\hat \Phi_\omega^{``out"}$ onto $\hat \Phi_\omega^{``in"}$ using 
(\ref{outmodepast}), and the definition of the inner product on ${\cal J}^-$, which is the complete Cauchy surface of the ``in" region. 
In this way one avoids having to deal directly with the subtleties of the horizon and the complications from curvature in the future region.
Using \cite{patatino}
\be
\langle \Psi , \Phi \rangle = i \int dv r^2 d\Omega \Bigl[\Psi^* (\partial_v \Phi ) - (\partial_v \Psi^*) \Phi \Bigr] \, ,
\label{innerpr}
\ee
and writing mode expansion $\Phi_{\omega}^{``out"}{}_{a} = \int_{\omega',b} \Bigl\{ \alpha_{ab}(\omega,\omega') \Phi_{\omega'}^{``in"}{}_{b}  
+ \beta_{ab}(\omega,\omega') \Phi_{\omega'}^{``in"}{}_{b}{}^{*}\Bigr\} \, $,
where $a,b$ designate outgoing ($o$) and ingoing ($i$) modes, and $\int$ designates a summation over discrete parameters and integration over
continuous ones, we can evaluate the $\alpha$ and $\beta$ coefficients. The completeness and orthogonality of the ``in" modes then implies that the ``in" operators evolve to ``out" operators according to
\be
\hat \Phi_{\omega'}^{``in"}{}_{b} = \int_{\omega,a} \Bigl\{ \alpha_{ab}(\omega,\omega') \hat \Phi_{\omega}^{``out"}{}_{a} 
+ \beta_{ab}{}^*(\omega,\omega') \hat \Phi_{\omega}^{``out"}{}_{a}{}^\dagger \Bigr\} \, .
\label{opexpansion}
\ee
Now, {\it imposing} the condition that the ``out" modes are also a complete set of modes, as well as orthogonal\footnote{This means, 
that the map between ``in" and ``out" regions is perturbatively {\it unitary}, and that we have

\centerline{$\int_{\omega, a} \Bigl(\alpha_{ab}(\omega, \omega') \alpha_{ac}{}^*(\omega, \omega'') 
- \beta_{ab}(\omega,\omega') \beta_{ac}{}^*(\omega,\omega'')\Bigr) = \delta_{bc} \delta(\omega'-\omega'') \, ,$}
\centerline{$\int_{\omega, a} \Bigl(\alpha_{ab}(\omega, \omega') \beta_{ac}{}^*(\omega, \omega'') 
- \alpha_{ab}(\omega,\omega') \beta_{ac}{}^*(\omega,\omega'')\Bigr) = \delta_{bc} \delta(\omega'-\omega'') \, .$}

and their transposes with respect to the matrix indices $a, b, c$.}
we can invert (\ref{opexpansion}) to find
\be
\hat \Phi_{\omega}^{``out"}{}_{a} = \int_{\omega',b} \Bigl\{ \alpha_{ab}{}^*(\omega,\omega') \hat \Phi_{\omega'}^{``in"}{}_{b} 
- \beta_{ab}{}^*(\omega,\omega') \,  \hat \Phi_{\omega'}^{``in"}{}_{b}{}^\dagger \Bigr\} \, .
\label{opexpansion2}
\ee
Using this expansion, one can compute the expectation value of the occupation number operator $N^{``out"}_a = \int_{\omega} 
\Phi_{\omega}^{``out"}{}_{a}{}^\dagger \Phi_{\omega}^{``out"}{}_{a}$ in the state $| 0_{in} \rangle$. Since
$\langle N^{``out"}_a \rangle = \int_\omega ||  \Phi_{\omega}^{``out"}{}_{a} |0_{in} \rangle ||^2 = \int_{\omega} N^{``out"}_a(\omega)$, 
from (\ref{opexpansion2}) and the definition of
$|0_{in} \rangle$ one finds $N^{``out"}_a(\omega) = \int_{\omega', b} |\beta_{ab}(\omega,\omega')|^2$.  

The black hole radiance at ${\cal J}^+$ is given by the expectation value of the $N^{``out"}_{outgoing}$.
Setting $a = o$, using (\ref{innerpr}) and the mode expansion, dropping the $u$-dependent pieces in the limit $u \rightarrow - \infty$ (which means, taking the modes
to ${\cal J}^-$ and using decoupling, or more formally, the Riemann-Lebesgue lemma), the coefficients determining the overlap 
of the ``out" outgoing modes with the ``in" modes are, in terms of the variable $x = v_H - v$,
\be
\alpha_{o,i}(\omega,\omega') = - \frac{1}{2\pi} \sqrt{\frac{\omega}{\omega'}} \frac{e^{-i(\omega-\omega') v_H} }{(2r_0)^{2ir_0 \omega'}}
\int_0^\infty dx e^{-i\omega x} x^{2i r_0 \omega'} \, , 
\label{bogoliubov}
\ee
and $\beta_{o,i}(\omega,\omega') = i \alpha_{o,i}(-\omega,\omega')$. To regulate the integral, one shifts the frequencies by 
$\pm i \omega \rightarrow \pm i \omega - \epsilon$, to ensure that the interactions are shut off in the infinite future and past.
Once this is done, (\ref{bogoliubov}) can be evaluated in terms of $\Gamma$-function as \cite{patatino}
\be
\alpha_{o,i}(\omega,\omega') = - \frac{1}{2\pi} \sqrt{\frac{\omega}{\omega'}} \frac{e^{-i(\omega-\omega') v_H} }{(2r_0)^{2ir_0 \omega'}}
\frac{\Gamma(1+2ir_0 \omega')}{(\epsilon + i\omega)^{1+2ir_0\omega'}} \, .
\label{bogoliubov2}
\ee
From this it immediately follows that  
$\beta_{o,i}(\omega,\omega') =  - i e^{-2\pi r_0 \omega' -2i\omega v_H} \alpha_{o,i}(\omega,\omega')$, 
where one uses $\ln(\epsilon + i\omega) = \ln(\epsilon - i\omega) + i\pi$ for the correct analytical continuation, that avoids branch changing.
Using a sum rule that follows from footnote 2, setting $\omega'' = \omega'$ and integrating the transpose of 
the first of the equations in footnote 2, one obtains 
$\int_{\omega, \omega', b} \Bigl( |\alpha_{ab}(\omega, \omega')|^2 - |\beta_{ab}(\omega, \omega')|^2\Bigr) = 1$. Substituting $\beta_{o,i}(\omega,\omega') =  - i e^{-2\pi r_0 \omega' -2i\omega v_H} \alpha_{o,i}(\omega,\omega')$ in this
formula yields $\int_{\omega,\omega',i} |\beta_{o,i}(\omega,\omega')|^2 = \int_{\omega} \frac{1}{e^{4\pi r_0 \omega} - 1}$, which implies that
{\it in the near-horizon limit},
the ``in" vacuum $|0_{in} \rangle$ in the outgoing regime is a state with particles in it, following the black body distribution 
\be
N_{outgoing}^{``out"}(\omega) = \frac{1}{e^{\omega/T_{BH}} - 1} \, ,
\label{occupationnum}
\ee
with the black hole temperature $T_{BH} = (4\pi r_0)^{-1}$ \cite{hawking,unruh}. This is depicted in Figure 2.
As we noted, this applies to any and all 
angular momenta in the mode expansion of $\nabla^2 \Phi = 0$. This would remain true {\it even} if the scalar field were massive. In the Schwarzschild region, the mass term would appear as an extra contribution in the radial equation $\propto m^2 (1-r_0/r) \phi_\omega$, which is clearly negligible as $r \rightarrow r_0$. In other words, a quantum field theory near the horizon becomes ultrarelativistic, and locally the asymptotic mass terms become irrelevant. Thus all the degrees of freedom thermalize with the black hole near the horizon.

Once we have determined the heat near horizon, in the state which started as the infalling Minkowski vacuum, we can find what of it channels the black hole away. Not all of this heat can escape from near the horizon to infinity, as depicted in Figure 3. The near-horizon heat
is squeezed in the region of space close to the future event horizon of the black hole, and to escape it must pass out of the potential barrier at $ r \simeq r_0$. 
Of all the modes,
\begin{figure}[hbt]
\begin{center}
\includegraphics[width=0.55\textwidth]{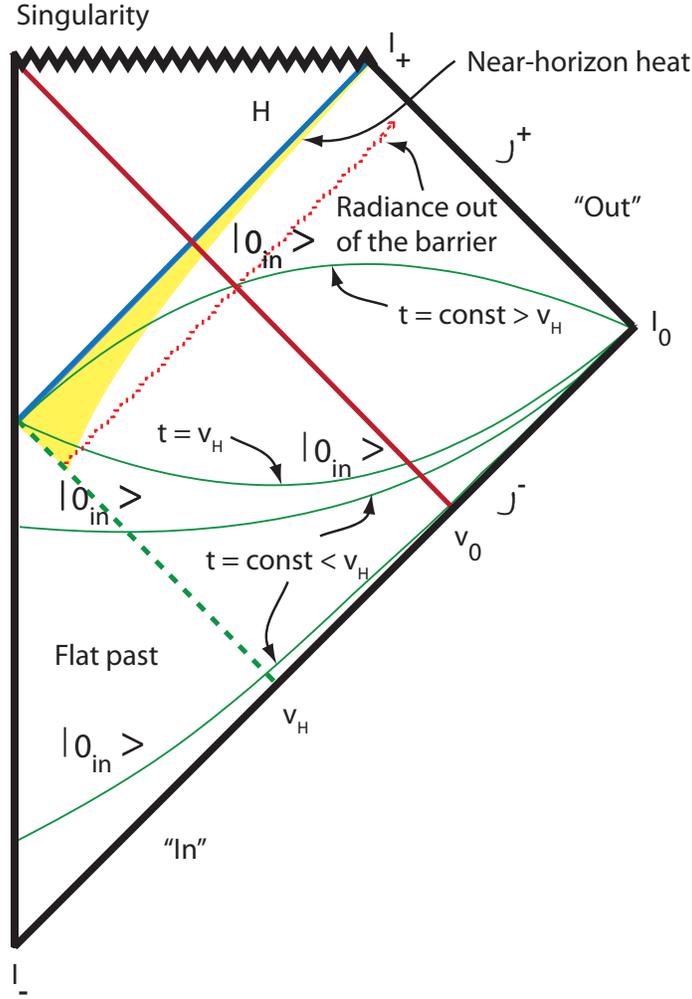}
\caption{Black hole heat}
\label{three}
\end{center}
\end{figure}
only the $s$-wave is relatively unsuppressed, since the barrier separating it from infinity is the lowest. The $l \ne 0$ modes are confined close to the black hole by the rapidly growing grey-body factors, which have been computed using black hole absorption by \cite{starobinsky,page} (for a review, see \cite{patatino}). 
The total black hole luminosity, which controls the black hole mass loss rate, is
\be
{\cal L}_{hawking} = \sum_{l=0}^\infty \frac{2l+1}{2\pi} \Gamma_{\omega,l} \int^\infty_0 d\omega \frac{\hbar \omega}{e^{\omega/T_{BH}} -1} \, ,
\label{luminosity}
\ee
and, using the grey-body factors \cite{patatino}, one finds that the 
main component of the escaping radiance outside of the barrier is the $s$-wave, i.e. a {\it single} mode, carrying out around $90\%$ of 
black hole radiance. 
If in addition the field is massive, even the $s$-wave will be prevented from escaping by the mass barrier \cite{mass}. Only the high frequency modes, with sufficient energy, can escape\footnote{This can be readily seen from considering radial geodesic equation, 
$(dr/d\lambda)^2 = E^2 - m^2(1-r_0/r)$, implying that all `particles' with energies $E<m$ cannot escape to infinity, but will fall back into the black hole after a finite excursion away from it \cite{mass}.}, meaning that the emission of modes with masses larger than the black hole temperature is exponentially suppressed. 

\section{Where Entropy Roams}

Since the black hole emits thermal radiation, it must have entropy. It was conjectured by Bekenstein \cite{bekenstein} that black hole's entropy is given by its area in
Planck units, $S = \frac14 M_{Pl}^2 A$, even before Hawking's discovery of black hole radiation. The presence of Hawking radiation confirms it 
easily, using the second law of thermodynamics: entropy density in the outgoing thermal radiation in the rest frame of a distant observer
is ${\cal S} \sim T_{BH}^3$, and the entropy transferred out per unit time is $\dot S \sim T_{BH}^3 A$. Hence the total entropy lost by the 
radiating black hole is $S \sim \tau_{BH} \dot S \sim M_{Pl}^2 A$, where $\tau_{BH} \sim M/(T^4_{BH} A)$ is the Hawking black hole lifetime.
In the simple field theory model, this entropy should count the number of modes in the heat bath near the black hole horizon. These are represented by all the $l=0$ and $l\ne 0$ modes in the $|0_{in} \rangle$ state, following the distribution (\ref{occupationnum}). 

To reproduce the entropy scaling with the area, when the exterior field theory is Lorentz invariant, and without mass gap so that the state 
{\it is} thermal  over the full range of frequencies,
we can integrate over the average number of particles of all spins in it, which is ${\cal S} \sim T^3$. Here $T$ is the black hole temperature measured by a {\it local} observer residing at some $r = {\rm const}$ outside of black hole, 
including the blueshift factor, $T = T_{BH}/\sqrt{g_{00}(r)}$.  The warping of the spatial surfaces $t = {\rm const}$ due to the black hole gravity must also be included since it sets the proper volume in which the entropy is stored. The covariant spatial volume element is 
$dV = \sqrt{g_{rr}} dr r^2 d\Omega = 4\pi \sqrt{g_{rr}} dr r^2$ where spherical symmetry allows us to ignore angular integrations. Hence the entropy in the state $|0_{in} \rangle$ on spatial surfaces 
$t = {\rm const} \gg v_H$ is, using $g_{rr} = g_{00}^{-1} = (1-r_0/r)^{-1}$,
\be
S \sim \int dV {\cal S} \sim T_{BH}^3 \int dr \, \frac{r^4}{(r-r_0)^2} \, .
\label{entropy}
\ee

However: this is divergent both in the UV and in the IR, and needs to be regulated. The IR cutoff is straightforward, setting it at
a distance $r_{IR} \sim r_0$, where the integral (\ref{entropy}) is $\sim T_{BH}^3 r_0^3 \sim {\cal O}(1)$.  
The UV cutoff is subtler. The way to pick it is to regulate the divergences in (\ref{entropy}) such that the resulting finite integral matches the Bekenstein-Hawking
entropy formula, $S \sim M_{Pl}^2 A$. This is the definition of the stretched horizon of \cite{lenny}, and quantitatively agrees with the definition
of the brick wall regulator of \cite{thooft}. So, integrate (\ref{entropy}) to some distance $r_{UV}$, where 
$S(r_{UV}) \sim M_{Pl}^2 A$; since $|r_* - r_0| \ll r_0$, the leading divergence is
\be
S  \sim T_{BH}^3 \int_{r_{UV}} dr \, \frac{r^4}{(r-r_0)^2} \sim \frac{r_0}{r_{UV} -r_0} \, ,
\label{entropy2}
\ee
using $T_{BH} \sim 1/r_0$ and ignoring ${\cal O}(1)$ factors. Then picking $r_{UV} \simeq r_0 + (M_{Pl}^2 r_0)^{-1}$ ensures that
(\ref{entropy2}) scales as the area in Planck units. Since the proper distance is $\ell = \int dr \sqrt{g_{rr}}$, this means that the
stretched horizon is a timelike surface at the proper distance $\ell \simeq \int_{r_0}^{r_{UV}} \frac{dr}{\sqrt{1-r_0/r}}$ from the horizon, or at
\be
\ell_{SH} \simeq \ell_{Pl} \, .
\label{properdist}
\ee
At this surface, the blueshifted black hole temperature, measured by the accelerated observer $\ell_{SH}$ away
from the horizon, is $T_{SH} = T_{BH}/\sqrt{1-r_0/r_{UV}} \simeq M_{Pl}$, using the above formula for $r_{UV}$. 
So an effective quantum field theory in the state $|0_{in} \rangle$, with most excitations in a heat bath at the temperature measured at 
infinity $T_{BH}$, is very hot locally with the Planckian 
temperature measured on the radiating surface $\ell_{SH}$. The escaping radiation redshifts to the asymptotic temperature $T_{BH}$ at infinity as 
it works its way out of the black hole's potential well. Defined in this way, the stretched horizon is very close to the black hole horizon, well inside the centrifugal barrier 
$V_{BH}$ in Eq. (\ref{radial}), which is peaked at $r = 3r_0/2$. 

This is a consistent low energy effective field theory representation of the
dynamical black hole as viewed by outside observers.
For example, in 't Hooft's original brick wall calculation
\cite{thooft}, the black hole entropy is given in terms of the free energy as $S = - \partial F/\partial T_{BH}$, where
\be
F =  - \frac{1}{\pi} \int^{\infty}_0 d\omega \int^{r_{IR}}_{r_{UV}} \frac{dr}{1-r_0/r} \int_0^{l_{max}(\omega)} \frac{dl (2l+1)}{e^{\omega/T_{BH}} -1}  
\sqrt{\omega^2 - (1-\frac{r_0}{r}) \frac{l(l+1)}{r^2}} \, ,
\label{thooft}
\ee
where $l_{max}(\omega)$ is defined as the largest angular momentum for which the discriminant of the square root is still positive.
Using new variable $y = l(l+1) (r-r_0)/r^3$, one immediately finds the dominant contributions to $F$, 
scaling as\footnote{A key feature of this calculation
is that the spectrum is Planckian over the whole range of frequencies between the IR and UV cutoffs.}
 $F \simeq - T_{BH}^4  \int^{r_{IR}}_{r_{UV}} dr \frac{r^4}{(r-r_0)^2}$, which yields precisely our heuristic formula (\ref{entropy2}).
 All the modes, $l=0$ and $l \ne 0$ contribute here, since this is the total free energy at a given time, rather than the energy 
 (and entropy) that goes out.  
This is further corroborated by the covariant computation of \cite{myers}. 
In fact, the picture which emerges from \cite{myers} is that the entropy estimate (\ref{entropy2}), that formally diverges as the cutoff is lifted,
has the same divergence as the Planck scale, such that when the Planck scale is additively renormalized entropy remains finite.
The regulated model of this black hole geometry is depicted in Figure 4. 
\begin{figure}[hbt]
\begin{center}
\includegraphics[width=0.55\textwidth]{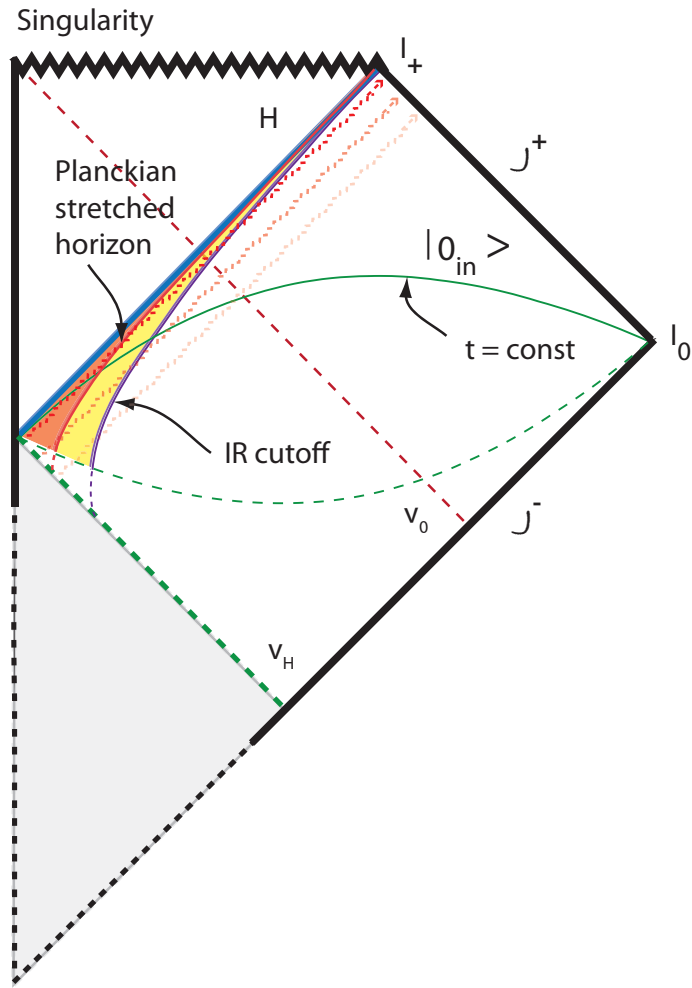}
\caption{Cutoffs and quantum states}
\label{four}
\end{center}
\end{figure}

Effectively, the spacetime sliver between the Planckian stretched horizon and the true event horizon is excised, ending on a hot surface where all local processes occur at near-Planckian energies. This region removed is rendered irrelevant by the Planck scale renormalization. Indeed, if one takes the quantum field theory cutoff at the Planck scale and regulates it 
with a Lorentz-invariant regulator, one finds that the divergence  in (\ref{entropy2}) has precisely the same leading order behavior as the one-loop counterterm renormalizing the Planck scale in flat space field theory 
\cite{myers}. Dropping higher frequencies and their entropy contributions simply means that their (infinite) contribution to the entropy is precisely compensated away by the (infinite) renormalization of Planck scale. Moreover, their omission does not change the radiance rate because on shell these modes would have a huge spin, and so are extremely strongly suppressed by the centrifugal barrier. Clearly the key for deducing this, and determining the radiation rates with the Planck cutoff is that the state they are in is $|0_{in} \rangle$.  This follows from Lorentz symmetry and the existence of the decoupling limit. In fact, this result is quite robust: more detailed investigations suggest that the low energy properties of the black hole radiance and entropic contents remain largely independent from the `trans-planckian' effects that are modeled by the rigid cutoff \cite{transplanckian}, 
{\it as long as they are computed in the state} $|0_{in} \rangle$, even if one considers UV corrections to the field theory which violate Lorentz symmetry but do not introduce explicit instabilities \cite{transplanckian}. 
The `cost' of this picture is that the microphysical details of black hole entropy accounting are delegated to the UV effects which are not under direct calculational control in the low energy theory below the cutoff, which however `knows" the correct result thanks to the divergences controlled by the symmetry. Traced back in time, all the Hawking quanta reaching the future null infinity ${\cal J}^+$ and taking away black hole's mass and entropy originate from somewhere in there. While the process of their creation is obscured from effective field theory by the stretched horizon, the accounting of modes is properly replicated, since in this case the only obstruction they encounter on the way out are centrifugal and mass barriers, well below the cutoff.

\section{And Then There Were Many}

When the field theory has a large number of modes in the IR the story becomes more complicated, since describing how black hole entropy is stored and radiated is exacerbated by the presence
of many modes (and possibly their interactions) \cite{species}. Nevertheless, we would expect the theory to obey the
standard laws of thermodynamics in the IR. Even if the field theory modes are strongly coupled, their thermodynamic potentials, and specifically 
internal energy and entropy, should scale with the temperature in the usual way, even if the numerical coefficients in these relations may be
corrected by the interactions. If the theory is heated up to some finite temperature, all the modes which are light at that temperature will
be excited, and their frequency distribution ought to follow a black body spectrum. 

In a black hole background, as long as gravity is universal, and all the IR modes are accessible to it, this suggests
that all the light IR modes will be thermally excited, and so contribute to the black hole entropy, and perhaps even radiance. Based on this simple set of assumptions, to estimate entropy  one could use Eqs. (\ref{entropy}) and (\ref{entropy2}), with a prefactor of $N$ counting all the light fields in the IR. The leading divergence is
\be
S \sim N \frac{r_0}{r_{UV} - r_0} \, .
\label{Ntropy}
\ee
If a flat space UV cutoff of the theory is high, say at the Planck scale, and we try to use it in (\ref{Ntropy}) by taking the stretched horizon at
$\ell_{SH} \simeq \ell_{Pl}$ from the event horizon, this would scale as $S \sim N M_{Pl}^2 A$, overshooting the black hole entropy by
a factor of $N$. This is the {\it naive} species problem \cite{species}: many distinguishable light modes in the theory can yield much more entropy storage than given by the black hole area. 

The error in estimating entropy by $S \sim N M_{Pl}^2 A$
 is in ignoring gravity. The entropy counting based on (\ref{Ntropy}) pushed all the way to Planck scale completely ignores gravitational interactions of field theory modes, which in fact dominate in the UV \cite{thooftdominance}. The point is to estimate just what the cutoff should be for using the flat space field theory to describe a black hole. This `geometric' cutoff is again given by the 
stretched horizon, which as before is defined as the distance at which $S(r_{UV}) \sim M_{Pl}^2 A$, with $S$ given in (\ref{Ntropy}). Combining these
two equations yields $r_{UV} \simeq r_0 + N (M_{Pl}^2 r_0)^{-1}$, which is significantly farther from the event horizon than $r_{UV}$ for a single
species. The proper distance between the multispecies stretched horizon and the event horizon is now
\be
\ell_{SH} \simeq \sqrt{N} \ell_{Pl} \, .
\label{stretN}
\ee
This means, that the low energy field theory description of the black hole state $|0_{in} \rangle$, taken to be a thermal state of all the light IR modes, can be trusted only down to the distance $\sqrt{N} \ell_{Pl}$ from the horizon. 
The graphic description of this situation is given in Figure 5. 
\begin{figure}[hbt]
\begin{center}
\includegraphics[width=0.55\textwidth]{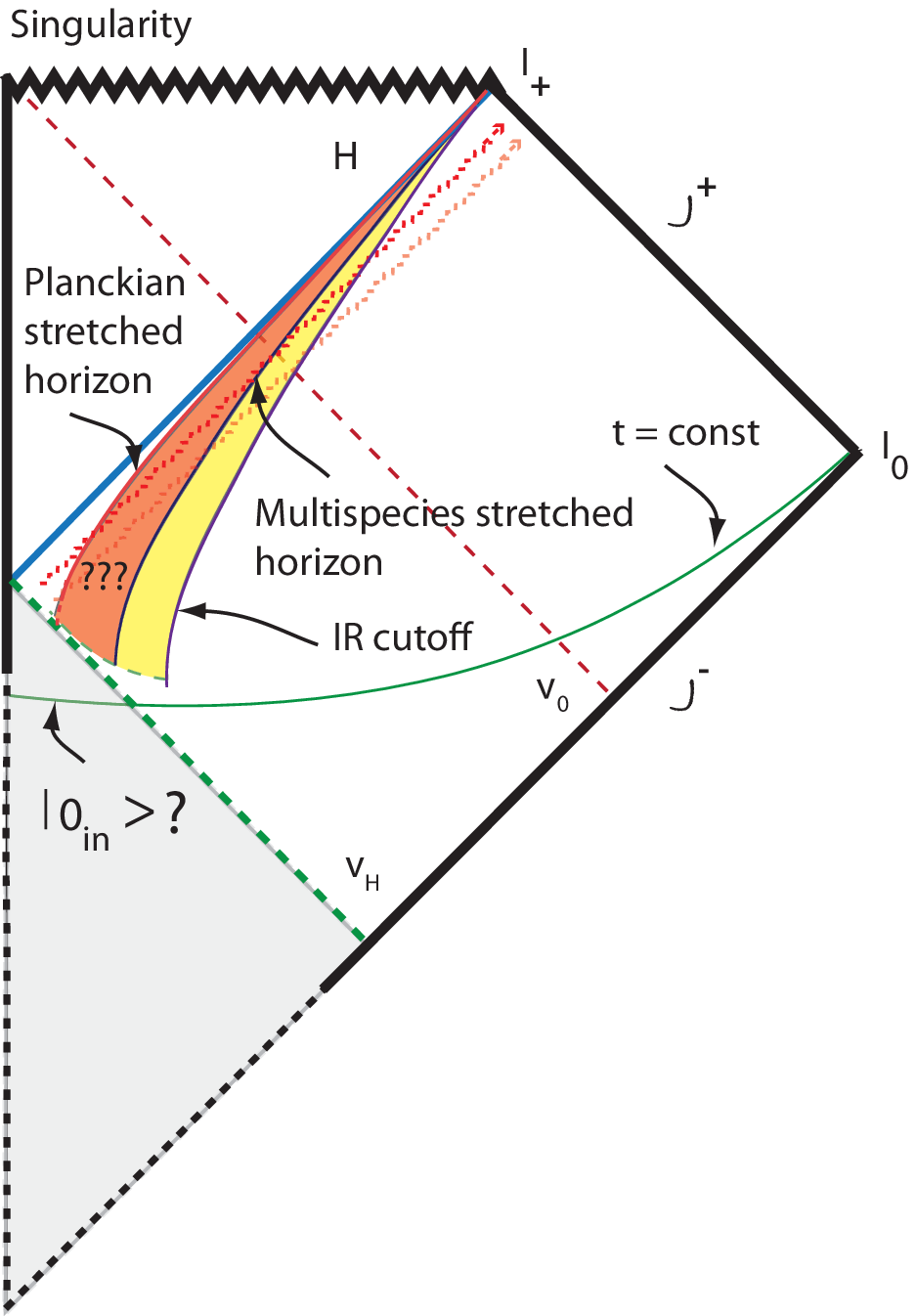}
\caption{Stretching the horizon with many species}
\label{five}
\end{center}
\end{figure}
So, even though the geometry seems to be perfectly well defined all the way to the horizon, the field theory defined by the low energy Lagrangian with many light species cannot be used to follow the presumed multispecies black hole thermal state into the region beyond the multispecies stretched horizon. This suffices to
reproduce the black hole entropy. But, any effects that can influence the radiation coming out from the horizon, which may occur in the region of geometry between the (near) horizon and the multispecies stretched horizon remain completely hidden. 

This situation is very uncomfortable. To define the thermal black hole state in the first place, and so to determine the populations of modes, both thermalized, and radiating to infinity\footnote{We stress that this is not the same, as we have seen in the case of angular momentum decomposition of a scalar field.} in the dynamical black hole setting (\ref{vaidya}) we have taken the theory all the way to the horizon in the ``in" vacuum. We have done it using the flat space field theory {\it before} the horizon formed, having no obstructions from neither the blueshifts nor the entropy counting issues. 
The features of this state could be verified by an ``out" observer receiving signals which originated from slices $t < v_H$, just before the horizon appeared. Since the local blueshifts render all the low energy scales irrelevant in the ``out" sector, the shell implosion forming the horizon only effects a Bogoliubov transformation on the 
various momentum modes to the leading order, rendering the ``in" vacuum into the thermal state in the ``out" region. With many species, and the cutoff of the ``out" region set up by the multispecies stretched horizon, this is {\it not} obvious any more: one cannot directly follow time evolution of the ``in" vacuum all the way to just before the horizon forms using the multispecies IR theory even in principle since the cutoffs jump on some of the $t = {\rm const}$ slices, as indicated on Figure 5. One will know entropy, assuming that the state $|0_{in} \rangle$ is thermal, but will not be able to verify the detailed interactions in this state beyond a certain time. The danger is that giving up the field theory description of the (trans) Planckian region close to the black hole undercuts the whole reasoning for having a thermal bath of quanta near the black hole, and makes the choice of the exterior quantum state of the black hole suspect. Alternatively, one may not be able to determine the backreaction of the field theory in the state $|0_{in} \rangle$, and may not be able to guarantee regularity of the horizon. 
As a consequence, one can't deduce the rate of energy loss from the black hole, since one doesn't have the full description of the theory near it.

If the UV cutoff of the low energy field theory is below the scale $\ell_{SH}^{-1}$, the whole point may be moot, since one can't directly use the low energy theory above such a low cutoff. An example would be trying to describe radiance of a black hole of an arbitrary mass solely in terms of pions and hadrons of some large-N QCD-like theory. If the black hole is very large and therefore cold, one could estimate the radiance rate by {\it assuming} 1) that the low energy theory remains in the thermal vacuum outside of the black hole even at energies much above the cutoff, which is not checkable in terms of pion and hadron mode functions, and 2) that the only grey body suppressions come from the usual centrifugal and mass barriers, which are at distances farther from both the stretched horizon and the UV cutoff of the low energy theory, and that there aren't any additional effects altering radiance rates between the stretched horizon and the (lower) UV cutoff. With this in place, one can take into account that at low energies the quark-gluon plasma undergoes a confining phase transition, implying that only QCD singlets escape to infinity\footnote{In \cite{lisa}, similar arguments were used to claim that black holes in RS2 braneworlds will not radiate only the few CFT singlets. We believe the real reasons for the suppression of CFT Hawking radiation are more involved, as we will argue in what follows.}. With these {\it assumptions} one will obtain the correct radiance in terms of pions of hadrons; but one can't be sure that the assumptions are precisely correct, since it may happen that at high energies above the cutoff (but below the Planck scale) the field theory interactions alter the radiance rate, yielding extra suppression. For a very small black hole, with the temperature above the confining phase transition this is not even viable because pions and hadrons do not exist anywhere in the space around it, and one would simply not be able to determine any grey body factors whatsoever. However, using the quark- and gluon-like states one would see that those are subject to the centrifugal and mass barriers, now much closer to the black hole -- above the confinement scale -- and check if there are any further suppressions from near-horizon field theory dynamics in the background geometry. This would settle any lingering doubt about there being any further suppressions of radiance from any dynamics below the scale of the stretched horizon and above the UV cutoff of the low energy theory of pions and hadrons\footnote{In fact there are claims that standard QCD interactions near black hole generate extra suppressions of Hawking radiance of hadrons \cite{heckler,cline}; however \cite{pagecarr} show that such extra suppressions are absent using QCD dynamics above the confinement scale.}.

If, on the other hand, the flat space UV cutoff of the theory is above the scale $M_{Pl}/\sqrt{N}$ this problem is harder, because the gravitational corrections to the low energy action may become crucially important near the cutoff. One needs a way of systematically including (possibly nonperturbative) gravitational effects in the field-theoretic description of the black hole. Arguments that the theory is fixed by a single UV scale are ambiguous in perturbation theory. The perturbative renormalization of Planck scale and entropy show \cite{myers} that the leading order divergences are the same, and can be subtracted away simultaneously, with the same counterterm. But this is not the same as demanding that the renormalized Planck scale is fixed by the finite contribution from the one-loop effects, which is a much stronger requirement. Nevertheless, the latter idea is appealing, and it is tempting to argue that the low energy field theory with gravitational corrections included really does not have as high a cutoff as without gravity, but actually ends at $M_{Pl}/\sqrt{N}$ 
\cite{inducedgr,gia}. 

Once one declares that the multispecies field theory must have a cutoff set by its stretched horizon, $\mu \sim M_{Pl}/\sqrt{N}$, in essence stating that gravity is induced \cite{inducedgr,induced}, and therefore `natural' , interesting possibilities for resolution open up. The main point is that when the theory is pushed up to the scale $\mu$, not only does the IR limit of the low energy field theory cease to be valid, but also the gravitational sector can soften up, with higher-order gravitational corrections becoming important. Yet, following the bottom-up approach, and simply extrapolating the IR theory, if it remains weakly coupled in perturbation theory at this scale, one still may not be a able say much.
However, if one {\it knows} what the UV completion actually {\it is}, one may just use it above the perturbative cutoff of the IR theory and push it to the black hole horizon, to see what really happens.

\section{Black Holes in the RS2 Braneworld}

The RS2 braneworld model is an example of how the UV completion, going beyond the cutoff of the IR CFT, fully resolves the species problem, and allows one to determine the Hawking radiance rate unambiguously. In RS2, the low energy quantum field theory is an $SU({\cal N})$ super-Yang-Mills CFT in strong coupling and large ${\cal N}$ limit, with many degrees of freedom $N \sim {\cal N}^2$. It is however cut off at a scale $\mu \sim L^{-1}$, where $L$ is the $AdS_5$ radius. This is where the Planck brane sits, inducing $4D$ gravity. Below the cutoff, the low energy CFT couples to $4D$ gravity 
\cite{rsholo,gubser}, whose $4D$ Planck scale $M_{Pl}$, the cutoff $\mu \sim L^{-1}$ and the number of degrees of freedom $N$ are related by $M^2_{Pl} \sim N \mu^2 = N/L^2$ \cite{induced}. Above the cutoff, one can define the UV completion of RS2 by using its gravity dual in $5D$ 
at distances shorter than $L$ (and resolving the Planck brane). Since the dual theory is weakly coupled bulk gravity (actually, weakly coupled string theory, strictly speaking) on $AdS_5 \times S^5$, it remains valid above the cutoff $\mu \sim L^{-1}$ with the only `difference' that the bulk modes at scales above the cutoff $\mu$ do not experience bulk warping. Instead,
at these scales and for high frequency modes, the bulk background behaves as a single flat extra dimension of size $ \sim L$. The propagating modes arrange into Kaluza-Klein sectors controlled by their behavior on the sphere $S^5$. Zero modes on $S^5$ are massless, while the modes that vary over $S^5$ become
massive in the $5D$ bulk, with their masses multiples of $\mu = L^{-1}$. On top of it, in addition to the zero-mode graviton, the theory may also couple to any (few) modes that might come in with the (resolved) cutoff brane (which are irrelevant for the issue of the black hole radiance into the CFT sector). The bulk description of the theory is weakly coupled and remains valid all the way to the bulk Planck scale, $\ell_{5} = M_{5}^{-1}$, beyond which one should use the full string theory in $10D$ \cite{rsholo,gubser}. Here, $M_5$ is related to $N$, $M_{Pl}$ and the cutoff $\mu = L^{-1}$ according to $N = M_{Pl}^2 L^2 = M_5^3 L^3$, so that $M_5^3 = M_{Pl}^2/L$.

Consider now a black hole in this setup, in an asymptotically flat space. 
From the bulk viewpoint, the black hole is highly asymmetrical, with the $4D$ horizon set by its mass, $r_0 = 2G_N M$, and its bulk extent warped down to $r_{bulk} \sim L \ln(r_0/L)$. The black hole is not stationary in the bulk, being accelerated toward the (removed) $AdS_5$ boundary, carried along by the (tensional) brane. The brane geometry is asymptotically flat. According to Maldacena AdS/CFT duality \cite{adscft}, one knows that the solutions of the classical bulk gravity equations describing the boundary geometry correspond to the configuration in the dual strongly coupled $4D$ theory which include {\it all} planar diagram corrections on top of the classical background. Specifically, they should also include Hawking radiation of all the modes available to the black hole, because of the universality of $4D$ gravity \cite{efk,takahiro}. 

The key issue is, which modes are available to the black hole to radiate away, and at which rates? Given that the CFT remains gapless all the way to infinity, with no confinement, one may think that all the $N$ degrees of freedom of the CFT can be radiated by a black hole sitting on the RS2 brane. This is supported by the computation of the corrections to the Newtonian potential at large distances between two probe particles placed on the brane, yielding
\be
\Delta V \sim - G_N mM L^2/r^3 \simeq  \frac{N}{M_{Pl}^2 r^2} V_N \, ,
\label{delV}
\ee 
with $N \simeq M_{Pl}^2 L^2 \simeq M_5^3 L^3$, both from the bulk \cite{RS2,gartan,giddings} and the dual $4D$ CFT point of view \cite{gubser,duff}. From the $4D$ point of view the corrections arise as the graviton vacuum polarization diagrams with the CFT in the loop, with the insertion of $\langle T_{\mu\nu} T_{\lambda\sigma} \rangle$. This precisely counts the number of CFT modes \cite{gubser}, independently of the UV cutoff at $\mu \sim L^{-1}$. There is no Yukawa suppression in this formula to indicate that any of the modes are suppressed by their incipient 4D masses. However, this formula also shows that the rate at which the CFT modes transfer momentum to generate the force is significantly weaker, yielding a potential that drops off as $1/r^3$ rather than the stiffer potential $1/r$ coming from the hard zero-mode graviton. This gives a clue, that the low energy CFT modes are not very good for transferring energy-momentum in 4D scattering processes.

What happens in a black hole background? It would be well-nigh impossible to compute the Hawking radiance in the same way as in weakly coupled theory as in our review above, because of strong couplings. Further, since the low energy CFT has a low cutoff, we wouldn't even know how to define the `infalling' vacuum of the theory at energies above the cutoff. However, one does expect that the theory follows standard laws of thermodynamics with all the states available to the black hole thermally excited, strong couplings or not. One may expect corrections to the coefficients appearing in the definition of thermodynamic potentials as the functions of extensive parameters, but overall parametric dependence should remain the same. Yet, given the many IR species, one knows that the regime of validity of the CFT description outside of black hole is limited by the multispecies stretched horizon. 

So let us for a moment assume that the calculated IR corrections to the Newtonian potential imply that all the CFT modes in the IR are accessible to the black hole. The total entropy in the black hole quantum state $| 0_{in} \rangle$ would have been $S \sim N T^3_{BH} \int_{r_{UV}} dr {r^4}/{(r-r_0)^2} \sim N {r_0}/{(r_{UV} - r_0)} $, as we disucussed in Eqs. (\ref{entropy}), (\ref{entropy2}) and (\ref{Ntropy}) above, with the stretched horizon is at the distance 
\be
\ell_{SH} \sim \sqrt{N} \ell_{Pl} = L = \mu^{-1}\, ,
\label{rs2sh}
\ee
from the event horizon, which is  precisely  at the RS2 CFT cutoff. At this distance from the black hole, the local temperature of the 
CFT in thermally excited state set by the black hole, in the units of the distant
observer, is $T = T_0/\sqrt{g_{00}(r_S)} \sim (\sqrt{N} \ell_{Pl})^{-1} = \mu$, following from (\ref{vaidya}) and
the scaling (\ref{rs2sh}). Closer to the black hole we cannot use the IR CFT; it doesn't exist. Moreover, since at the stretched horizon the temperature $\sim \mu$ is in fact the blueshifted Hawking temperature at infinity, and since most of the Hawking radiation at infinity arrives from the energy bin around it's temperature, we couldn't even be sure the spectrum would be exactly thermal if we aren't able to say what happens with the modes with energies much higher than $\mu$ at the stretched horizon. Many of the modes needed to take the energy to infinity simply aren't there. So to make certain what happens in the sliver between the Planckian and the multispecies stretched horizon, another description of the quantum field theory is needed. There, we can use the UV completion of the RS2 setup.

\begin{figure}[hbt]
\begin{center}
\includegraphics[width=0.55\textwidth]{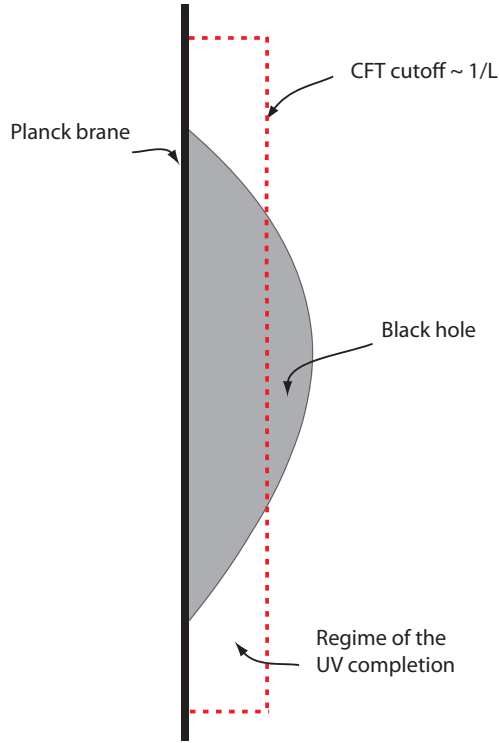}
\caption{Diving near the black hole horizon in RS2}
\label{six}
\end{center}
\end{figure}

Closer to the black hole the theory reduces to weakly coupled bulk gravity on $AdS_5 \times S^5$ with the extra normalizable $4D$ graviton zero mode. The bulk background behaves as a single flat extra dimension of size $ \sim L$. Because we wish to describe large black holes on the RS2 brane, with the $4D$ event horizon
$r_0 = 2G_N M \gg L$, the horizon in the bulk extends to $r_{bulk} \simeq L \ln(r_0/L) > L$, which we can take to be significantly larger than the AdS radius. Thus we can treat the theory as a $4D$ one with towers of KK modes, with, in general, gapped continuum spectra. The   
bulk fields are classified by their behavior on $S^5$. All the fields which vary over $S^5$
are massive in the $5D$, with the mass gap set by $\mu = L^{-1}$. In the effective $4D$ language, these fields come as continuum spectra with mass gap also set by $L^{-1}$, because the size of the effective fifth dimension is also $L$ (see Figure 6; there we could have depicted the bulk stretched horizon around the black hole, but it is irrelevant for the argument and we will ignore it because it is small, and the description is effectively $4D$ anyway), and with the continuum variable parameterized by the bulk momentum. None of these modes will be emitted as Hawking radiance because they are too heavy and large black holes are too cold, with $T_{BH} \sim 1/r_0 \ll L^{-1}$. 

That leaves only the zero modes on $S^5$ as the degrees of freedom possibly accessible to black hole to radiate away. They are massless in $5D$, but may be gapped continua in the effective $4D$ theory, if they have $5D$ momentum. 
If their bulk momenta are above the CFT cutoff $\mu = L^{-1}$, they are again too heavy to be emitted by the large black holes. On the other hand, those with $5D$ momentum below the black hole temperature $T_{BH}$ could in principle be emitted. 
But: these modes are heavily suppressed in the region of space surrounding the black hole at the proper distance $L$ from the horizon!

A simple way to understand how this suppression works is to introduce a regulator brane in the IR, at some very large proper distance in the bulk, 
$R \gg L$. So long as the regulator is much farther in the bulk than the bulk extent of the black hole horizon, it is completely irrelevant for the black hole.  While it gaps the bulk spectra, the mass gap it induces is
$m_{gap} \sim L^{-1} e^{-R/L}$, and if the IR brane is far down the $AdS$-throat, $T_{BH} \gg m_{gap}$, and so this does not introduce any significant suppressions to the radiance\footnote{This is in contrast to how the regulator brane in the IR was used in \cite{lisa}, where the black hole's horizon in the bulk extended beyond the regulator. In our case the regulator brane is merely a spectator.}. The modes which could be emitted are $4D$ $s$-waves with bulk momenta, or therefore effective $4D$ masses\footnote{Near the black hole we can take the space to be locally flat, with $p_4^2 = p_5^2 = m^2$, due to blueshift.}  $m$ which are 
smaller than $T_{BH}$. 
Then, since the masses of light states are $m \sim n m_{gap}$ \cite{smallnos}, the number of the states lighter than the black hole temperature 
per a bulk field can be huge,
\be
N_{light} \simeq \frac{T_{BH}}{m_{gap}} \simeq T_{BH} L e^{R/L} \, .
\label{nostates}
\ee
In the continuum limit $R \rightarrow \infty$ this number saturates at $N \la M_{Pl}^2 L^2$.  So a large fraction of the low energy CFT spectrum is in principle available to be radiated away.

However, the light modes are suppressed by the bulk potential barrier, i.e. the volcano of RS2. Their radial behavior is controlled by the solutions of the Schr\"odinger equation\footnote{Here as noted above $m^2$ is the eigenvalue of the $4D$ Klein-Gordon operator $\nabla^2_4$, from the assumption that we can think of the black hole locally as a section of a warped Schwarzschild string; while this is not quite true, it will serve our purpose.}
\be
\frac{d^2 \psi}{dz^2} + (m^2 - \frac{\kappa}{(z+L)^2}) \psi = \sigma \delta(z) \psi \, , 
\label{schrodinger}
\ee
with appropriate boundary conditions on the UV brane at $z=0$ and the IR brane at $z_{IR} = L e^{R/L}$, with $z_{IR}$ being the conformal distance in the bulk
(see, eg \cite{smallnos} for details). The parameter $\kappa$ depends on the specific details of the field theory sector in which $\psi$ originates, such as spin, bulk mass etc, and needs to obey $1+ 4\kappa \ge 0$, to ensure the absence of instabilities \cite{bf}. We can easily understand the behavior of these wavefunctions without solving the equation. For light modes $m \ll L^{-1}$, the potential is only relevant in the region $z \sim m^{-1}$ around the UV brane. Farther away, it plays no role, and the mode propagates as a free wave out to the IR brane at $z_{IR} = L e^{R/L}$. Thus, the mass gap and the wavefunction normalization are controlled by $z_{IR}$, being $z^{-1}_{IR} \ll r_0^{-1}$ and $1/\sqrt{z_{IR}}$, respectively. On the other hand, the potential suppresses the wavefunction near the UV brane. To find the suppression, note from (\ref{schrodinger}) that the wavefunction $\psi$ depends on $m$ and $z$ only through the combination 
$m(z+L)$. Near $z=0$, for masses $m < T_{BH} \ll L^{-1}$, we can neglect $m^2$ in (\ref{schrodinger}), to find that the solutions behave as
\be
\psi = \bigl\{[m(z+L)]^{p_+} + {\cal O}(1) (mL)^{\gamma} [m(z+L)]^{p_-}\bigr\}/\sqrt{z_{IR}} \, 
\label{psi}
\ee
where $\gamma$ is related to the powers 
$p_\pm = (1\pm\sqrt{1+4\kappa})/2$, 
being $\gamma = p_+-p_-$, or, for the spin-2 case with a bound state, $p_+=5/2$, $p_- = -3/2$ and 
$\gamma = 2$. 
Hence in the region of space at distances $L$ from the horizon of the black hole, these wavefunctions have very little support. They are $\psi \simeq (mL)^{p_+}/\sqrt{z_{IR}}$ or, in the case of bulk gravitons,
$\psi \simeq (mL)^{1/2}/\sqrt{z_{IR}}$ \cite{smallnos}. The probability of having these states near the black hole, in the `box' of size $L$ near the UV brane (as depicted in Figure 6) is $\sim L |\psi|^2 \la (mL)^{2p}L/z_{IR}$,
where $p = p_+$, or $1/2$ for gravitons. Further, among all of the $4D$ modes of these states, only the $s$-waves can contribute to black hole radiance. 
Thus the total contribution of these states to Hawking radiation from the black hole cannot exceed
\be
N_{light} \times L |\psi|^2 \la (T_{BH} L)^{2p+1} \, ,
\label{probability}
\ee
and since $2p+1$ is at least $2 + \sqrt{1+4\kappa} \ge 2$, the lowest value of the power in (\ref{probability}) is {\it nonzero}. This means that the total outgoing radiation flux from the black hole is
\be
{\cal L}_{BH} \sim (T_{BH}L)^2 {\cal L}_{hawking} \sim (\frac{T_{BH}}{M_{Pl}})^2 N {\cal L}_{hawking} \, ,
\label{cftflux}
\ee
where ${\cal L}_{hawking}$ is the radiance rate for a single hard species of particle.
Because $T_{BH} L < 1$, it follows the total contribution of the light modes to Hawking radiation isn't even ${\cal O}(1)$, let alone the full number of IR states $N$\footnote{In the limit of small black holes, 
$r_0 \la L$, these formulas do not apply. For these black holes, bulk warping is negligible, and the enhanced spatial rotation symmetry implies that the emission is suppressed by the $5D$ centrifugal barrier as explained in \cite{robertoco}.}. The suppression is a consequence of the bulk warping and the potential barrier it induces\footnote{That bulk geometry may be suppressing the radiation has been stressed by Rob Myers; we thank him for sharing his insights.}. Note, that any dependence on the regulator brane cancelled out in (\ref{probability}), and so this persists even when CFT is completely gapless.

Let us clarify a subtlety in this discussion. 
We have been working in the UV regime, based on bulk gravity on $AdS_5 \times S^5$ which is the one valid microscopic 
description of the black hole state 
{\it inside} the box of size $L$ around it. We have combined the bulk mode suppressions determined by warping inside this box with the standard grey body factors induced by the centrifugal barrier at distances ${\cal O}(1) r_0$ from the black hole, to argue that the total suppression of the
radiance is large. At the barrier, we could have already resorted to the effective CFT description in the IR. Yet, since the CFT and the dual bulk description are {\it completely equivalent} below the cutoff, and since the theory looks completely $4D$ at distances $ \gg L$, we can continue to use the bulk description near and over the centrifugal barrier because it is the picture which allows us to get close to the black hole. Once past the barrier, we can translate the results into the CFT language, by using energy conservation, and determine the outgoing CFT flux of Eq. (\ref{cftflux}). The principal new element of this calculation is that besides the centrifugal and mass barriers, which for a large black hole are much below the cutoff, there is an additional suppression above the cutoff scale, which is hard to note directly with the low energy CFT, especially at high energies where the CFT doesn't even exist. 

At the same time, black hole entropy as computed by the external observer comes out right using the low energy CFT up to the multspecies stretched horizon at $\sqrt{N} \ell_{Pl} = L$. From the bulk point of view, the bulk modes contribute {\it en masse} to black hole entropy, 
and once very close to the horizon, one needs to resort to full $5D$ bulk gravity, with the $5D$ stretched horizon at $\ell_5$, to add them up. 
With this, the factor of $N$ is absorbed into the renormalization of the $4D$ Planck scale, $M_{Pl}^2 \sim  N \mu^2$, and so the
entropy scales as in Eq. (\ref{entropy2}). 
Indeed, since the black hole's horizon in the bulk does not extend past $r_{bulk} \sim L \ln(r_0/L)$, thanks to the warping of the bulk $AdS$ geometry
most of the horizon area is inside the `box' of the size of $L$ near the UV brane (see Fig. 6). The black hole area computed using $5D$ is 
$A_{bulk} \sim r_0^2 L$. Using $5D$ Bekenstein-Hawking entropy formula in this region gives $S_{BH} \sim A_{bulk} M_{5}^3 \sim r_0^2 L M_5^3$. However
since $M_{Pl}^2 \sim M_5^3 L$, this precisely implies that $S_{BH} \sim r_0^2 M_{Pl}^2 \sim A_{4D} M_{Pl}^2$, which means that the bulk modes produce the same answer for black hole's entropy as the low energy CFT.  This argument is very similar to the points made in \cite{robertoco} in the context of small black holes in ADD. A rather surprising consequence which follows from it, is that the RS2 black holes are in a sense never large, regardless of how far their horizon goes along the brane.

The suppression of black hole radiation rate of CFT modes follows from the combination of the low CFT cutoff $\mu \sim L^{-1} \ll M_{Pl}$, a large number of modes $N \gg 1$, and a weakly coupled bulk dual UV completion, with a built-in covariant mechanism for suppressing most Hawking quanta consisting of the centrifugal, mass, and bulk  barriers. The suppression is stronger than the advocated suppressions from strong coupling effect \cite{lisa,ruth,wiseguy}. 
The direct estimates of black hole radiance using low energy $4D$ CFT below the cutoff are just too naive, being complicated by both the strong coupling and, crucially, low cutoff which obscures the bulk barrier suppression. The strong coupling of the IR CFT, however, does play a role. It is essential for facilitating the existence of the weakly coupled bulk dual, which turns out to be a reliable probe of the black hole geometry all the way down to the horizon. It implies that the real field theory cutoff which one encounters in the black hole radiance description is in fact the string scale $\sim \ell_5$, beyond which one needs to use full 
$10D$ string theory rather than bulk gravity on $AdS_5 \times S^5$ to obtain microscopic description of black hole radiation. In this way, the theory automatically resolves the species problem, naturally supplanting the low energy limits for the UV extensions that can be systematically connected to the regime of full quantum gravity\footnote{With the disappointing, but sobering consequence that the problem of black hole information retrieval remains as elusive as ever.}. This resolution is a feature of the strong coupling in field theory, which facilitates the existence of the weakly coupled dual, and has nothing to do with strong gravitational couplings at high energies which will occur near any black hole horizon. For these effects one really needs to understand what happens at the Planck scale, but they will in general not have any consequences for low energy Hawking radiance as long as the low energy theory is a meaningful Lorentz-invariant quantum field theory, as we discussed at the end of Section 3. The issues discussed here are connected to the UV cutoff, and strong coupling, in field theory, and their location relative to $M_{Pl}$, and not about what happens to gravity at the Planck scale.

Are these conclusions universal? One may wonder if it might be possible to simultaneously resolve the species problem without suppressing black hole radiation rates.
As we saw, while the CFT picture of RS2 with a large number of CFT modes in the IR appeared intriguing, it didn't yield accelerated black hole decay due to the suppressions calculable in the weakly coupled bulk dual. The bulk dual, on the other hand, was the right setup because of the low CFT cutoff. If, on the other hand, there were a theory where the UV cutoff of the quantum field theory with many degrees of freedom (and many distinguishable $s$-wave modes) were much higher, significantly above the stretched horizon cutoff coming from the interplay with the gravitational sector, and where the theory didn't produce large suppressions of radiation rates, black hole decay might be faster. Since the multispecies stretched horizon is always at $\ell_{SH} \sim \sqrt{N} \ell_{Pl}$, and the local temperature of the black hole thermal state on it is $T = T_{BH}/\sqrt{g_{00}} 
\sim 1/\ell _{SH}$, requiring the field theory cutoff to be much higher, $\ell_{SH} \, \mu \gg 1$, implies that the $4D$ renormalized
Planck mass $M_{Pl~ren} \ni \sqrt{N} \mu$ is {\it much} lower than $\sqrt{N} \mu$. It should receive 
large negative contributions from some sectors of the theory:
\be
M_{Pl~ren}^2 = N \mu^2 - {\cal M}^2 \, .
\label{mplcorr}
\ee
In such a theory, the multispecies stretched horizon would satisfy $\ell_{SH} \sim \sqrt{N/(N \mu^2 - {\cal M}^2)}$, and so we
could have
\be
\ell_{SH} \, \mu \sim \Bigl(\frac{\mu^2}{\mu^2 - {\cal M}^2/N}\Bigr)^{1/2} \gg 1 \, ,
\label{strhorneg}
\ee
provided that the negative contribution ${\cal M}^2$ is sufficiently large.  A relatively simple framework to model such a behavior classically could be a variant of the DGP brane-induced gravity \cite{dgp} with $AdS$ bulk and {\it negative} brane-localized graviton terms, extending the setup of 
\cite{tanaka}. If some bulk modes were less suppressed near the brane, such a setup might give a way of enhancing black hole decay, that could be studied using the methods of bulk gravity. On the other hand, if such a framework turned out to be pathological, one could find further clues for learning how gravity affects UV behavior of field theories. Either way, it would be interesting to check it, and come up with more such examples. 

\section{Summary}

To summarize, the main conclusion of this paper is that knowing a low energy effective field theory only may not suffice to describe Hawking radiation from a black hole. One may need the UV completion of a low energy theory to really understand what happens. The reason is that the IR theory may 
have states which are composites of some of the UV modes which a black hole cannot efficiently radiate, because of suppressions based on their internal structure\footnote{Such issues were already encountered in studying how charge emission from the branes is suppressed in \cite{sergey}. We thank Sergey Dubovsky for pointing this out.}. Thus low energy theory may be blind to additional suppressions of black hole radiation rates generated by processes which operate {\it above} the UV cutoff of the low energy effective field theory, but below the Planck scale.

This implies that black holes in RS2 braneworlds will not radiate the CFT modes at the same rate as the hard brane modes. The radiation rates of CFT are strongly suppressed, being at most comparable to the rates of many fewer hard modes for very hot black holes. The point is that at distances closer than
$L \sim \sqrt{N} \ell_{Pl}$ low energy CFT doesn't even exist. Instead, since the CFT is strongly coupled, one can use its bulk dual, which can be formulated using AdS/CFT correspondence, that remains well behaved very close to the black hole. It shows that most bulk modes are not emitted because of the large gray body factors induced by the centrifugal, mass and bulk volcano barriers. This actually agrees with the perturbative calculations of long range forces due to the CFT corrections \cite{gartan}, where all $N$ CFT modes contribute, but much more weakly than the hard zero mode graviton. The strong coupling does play a role in the argument, albeit an indirect one, giving a way of defining the bulk dual. 
While solving bulk equations {\it does} include quantum corrections \cite{efk}, the radiative losses are very suppressed and they do not appear at the same level as the quantum corrections which maintain staticity. This is the correct physical interpretation of  
the numerical results obtained in \cite{wiseguy}, rather than imagining ambiguities in the quantum state of a strongly coupled CFT in a black hole background. It still remains to understand just what the corrections found numerically actually are (see \cite{marolf} for some new progress in this direction).
It would also be interesting to see how robust these conclusions are in other frameworks, where gravity is not induced.

\vskip.5cm

{\bf \noindent Acknowledgements}
 
\smallskip

We thank Sergey Dubovsky, Sandro Fabbri, Albion Lawrence, Markus Luty, Don Marolf, Matt Kleban, Norihiro Tanahashi, Takahiro Tanaka, Giovanni Villadoro and especially Roberto Emparan, Rob Myers and Lenny Susskind for many valuable discussions. We also thank Roberto Emparan, Matt Kleban, Albion Lawrence and Norihiro Tanahashi for useful comments on the manuscript. NK thanks the Benasque Center for Physics for kind hospitality during the inception of this work. The work of NK is supported in part by the DOE Grant DE-FG03-91ER40674.

\end{document}